# Transverse redshift effects without special relativity

**Eric Baird**  (eric_baird@compuserve.com)

Transverse redshift effects are sometimes presented as being unique to special relativity (the "transverse Doppler effect"). We argue that if the detector is aimed at 90 degrees in the laboratory frame, most theories will predict a redshifted frequency at the detector, although these predictions can be concealed by specifying that angles should be defined in a frame *other* than the laboratory frame. These redshifts are often stronger than special relativity's predictions. We list some of the situations in which lab-transverse redshifts would be expected.

## 1. Introduction

According to Einstein's special theory, light-signals coming from an object moving in the laboratory frame should have an increased wavelength when they arrive at a transversely-aimed lab-frame detector [1]. These "transverse redshifts" are sometimes presented as being a unique feature of special relativity, *e.g.* Rosser 1964:

> "… According to the theory of special relativity, if a beam of atoms which is emitting light is observed in a direction which according to the observer is at right angles to the direction of relative motion, then the frequency of the light should differ from the frequency the light would have if the source were at rest relative to the observer. This is the transverse Doppler effect. According to the classical ether theories there should be no change in frequency in this case." [2]

Other reference texts agree that transverse redshifts should not occur in classical theory [3][4], but are less specific about how the word "transverse" should be interpreted.

We show that Rosser's statement is incorrect, and that not only are "laboratory-transverse" redshift predictions common to a range of models, but that many of these predicted redshifts are stronger than their "special relativity" counterparts.

In this paper, we briefly look at and list the lab-transverse predictions of a number of different models.

## 2. "Stationary" and "moving" aether predictions

### "Aberration shift"

If we assume that light travels throughout space at *c* relative to the observed object, aberration effects cause an observer aiming their detector at 90° degrees in their own frame to see more of the "back-side" of the moving object, and can lead the observer to expecting to see a partial recession redshift [5] (*e.g.:* Lodge "*… Doppler effect caused by motion of the observer is … a case of common aberration.*" 1893 [6] "*.. a spurious or apparent Doppler effect.*" 1909 [7] ).

We can rederive these effects by starting with special relativity (which "relativises" the stationary-aether and moving-aether calculations) and working backwards to find the original moving-aether predictions.

### Non-transverse shift tests

Special relativity's transverse predictions are sometimes tested experimentally by measuring the non-transverse ("radial") frequency shift relationships, and then analysing the data to find a residual Lorentz component after first-order propagation effects have been accounted for [8]-[14].

The three main Doppler equations for the apparent frequency *f'* and apparent front-back depth *d'* of a receding or approaching radiating object [15][16][17], with *v* as recession velocity, are:

$$d'/d = f'/f = (c-v)/c \qquad \text{… (1)}$$

$$d'/d = f'/f = \sqrt{(c-v)/(c+v)} \qquad \text{… (2)}$$

$$d'/d = f'/f = c/(c+v) \qquad \text{… (3)}$$

Special relativity's "relativistic Doppler" predictions **(2)** are the root-product average of the predictions associated with "absolute aether"s that are **i)** stationary in the emitter's frame **(1)** and **ii)** stationary in the observer's frame **(3)** [18].

Any model that generates the first-order Doppler equation **(1)** should give a residual Lorentz-*squared* redshift when stationary-aether propagation effects **(3)** are divided out, a stronger result than special relativity's single residual Lorentz redshift.

In the case of Ives-Stilwell 1938 [8], the mean position of approach- and recession shifted spectral lines gives a central position that is not affected by velocity with **(3)**, and that has velocity-dependent positional offsets with **(1)** and **(2)**.

**Transverse motion**

The same relationships should hold for data taken at other angles in the laboratory frame – any "special relativity" result should be interpretable either as a stationary-aether propagation effect supplemented by a Lorentz redshift (time-dilation of the moving emitter), or as a moving-aether propagation shift supplemented by a Lorentz blueshift (time-dilation of the moving observer's reference-clocks).

Where the special theory predicts a lab-transverse Lorentz redshift, an unmodified "moving-aether" model should (again) predict a Lorentz-squared lab-transverse redshift (*Note:* all viewing angles must be specified in a particular frame to avoid aberration issues, otherwise this approach will fail [18] ).

### 3. Emitter-theory

If we are only observing a single object, the simplest predictions for a "ballistic light-corpuscle" model superimposed on flat spacetime should coincide with the predictions for an absolute aether moving with the object (lab-transverse Lorentz-squared redshift).

### 4. Dragged-light models

If light is completely dragged by a particle-cloud or object, we should again expect the most extreme scenario (where dragging is effectively absolute over extended regions of space), to be equivalent to a "moving aether" model, giving us a Lorentz-squared lab-transverse redshift.

Dragged-light models producing weaker dragging effects (or with more "democratic" dragging characteristics) should produce correspondingly weaker lab-transverse frequency changes.

### 5. Relativistic calculations using the emitter-theory shift equation

In another paper, we have derived the relativistic aberration and wavelength-changes associated with **(1), (2)** and **(3)** [19]. In that exercise, the relativistic application of the emitter-theory equation is once again associated with a Lorentz-squared "lab-transverse" redshift prediction.

In a round-trip version of the experiment (where a signal is aimed and the reflection received at 90° in the same frame), **(1)** gives a double Lorentz redshift and special relativity gives a null result [19].

### 6. Gravitational redshifts

Verifications of general relativity's gravity-shift predictions are sometimes used as indirect supporting evidence in favour of the special theory.

The prediction that light from high-gravity stars should be seen to be spectrally shifted was made by John Michell in 1783, and again by Einstein in 1910 [20]. If we calculate the strength of the effect by dropping an object across a gravitational gradient and using Doppler equation **(1)** to calculate its final motion shift (Einstein [21], MTW [22] §7.2), we get a one-way gravity-shift prediction of $\Delta E = \sim gh/c^2$ (good Earth-surface approximation), and $\Delta E = 2gh/c^2$ for round-trip shifts (exact relationship) [23].

Verifications of these relationships are often considered to be verifications of general relativity [24][25], although they do not depend on general relativity's mathematics, special relativity's frequency-shift relationships, or the principle of relativity.

### 7. Centrifugal redshifts

The equivalence principle requires that centrifugal redshifts must be calculable from gravitational principles [26], because of the apparent outward gravitational field seen in the rotating frame (the "Coriolis field" [27]).

If we attach two clocks to the centre and to the rim of a rotating disc, observers in the disc's rotating frame are entitled to claim that the disc is immersed in a effective gravitational field that pulls objects away from the rotation axis. We can then apply the general arguments given in Einstein's 1911 gravity-shift paper for signals passed through this field [21] to argue that the perimeter clock must run more slowly than the central clock.

These calculations do not require special relativity.

**Huyghens' principle and gravitation**

If two light-clocks *do* have a genuine measurable difference in clock-rate, we can apply Huyghens' principle to the apparent lightspeed differential between the two regions and predict a deflection of lightrays towards the slower clock [21][28]. By this argument, an effective gravitational field should be present in any experiment producing physical clock-rate differences.

ArXiv reference: http://xxx.lanl.gov/abs/ **physics/0010074**

## 8. Other rotating-body problems

Similar considerations apply to the Hafele-Keating experiment [29][30] and other experiments involving the comparison of rates of clocks orbiting with and against the earth's rotation (*e.g.* GPS and other satellite-based systems [31] §3 pp.54-64).

If a clock-rate difference is large enough to be deemed "significant", then the geometrical deviation from flat spacetime should be considered to be equally "significant" (since the former should be calculable from the latter).

### Einstein's equatorial clocks

The issue of gravitational-equivalence is nicely illustrated by the example in section 4 of the "electrodynamics" paper, in which Einstein suggests that a clock at the earth's equator should tick more slowly than one at the pole.

If we are using sea-level clocks, these gravitational effects conspire to make the effect disappear [32] – if a sea-level clock-rate differential is associated with a gravitational gradient, the earth's oceans should flow "downhill" across this time-dilation gradient towards the equator, only reaching equilibrium when all parts of the ocean surface have the same clock rate (the resulting equatorial bulge should, of course, also be calculable from more conventional "centrifugal force" arguments).

### More complex problems

Although it is useful to be able to calculate clock-lags by assuming flat spacetime and applying a Lorentz correction, the success of this approach over small regions does not mean that these are intrinsically flat-space problems (cartographers once used similar equations to compensate for the Earth's curvature, but their success did not prove that the Earth's surface "really was" flat). We would suggest that where "flat" and "gravitational" arguments disagree, the second approach may have greater validity.

The calculation of route-dependent gravitational effects from *apparent* clock rate differences is a much more complex subject, [33][34] and is beyond the scope of this paper.

## 9. Thermal redshifts

Similar arguments can be applied to the case of the thermal second order Doppler effect in $Fe^{57}$ [25][35]. If the $Fe^{57}$ atoms have "significant" velocities while locked into a "stationary" crystal lattice, then they must also be continually undergoing "significant" accelerations.

## 10. Muon lifetimes

"Muon-decay" experiments are widely cited in textbooks as supporting evidence of special relativity's time-dilation predictions [36].

C.M. Will [31] Appendix: pp.245-257:
> "But the [upper atmospheric] muon is so unstable that it would decay long before reaching sea level … if it weren't for the time dilation of special relativity, which increases its lifetime as a consequence of its high velocity."

This statement about the time-dilated muon depends on the assumption that the speed of light is "really" fixed in the observer's frame – but since the special theory ought to predict the same outcome when we assume that lightspeed is fixed in the *object's* frame, we are also entitled to claim, with equal validity, that the muon's ageing rate is anomalously fast, and that (with a fixed lightspeed in the muon frame) the muon would actually penetrate *further*, if is was not for the time-compaction effect of special relativity!

The "muon" statement obviously involves a certain amount of interpretation being applied to the experimental data. If we return for a moment to Newtonian mechanics, the muon's decay position $x$ for a given Newtonian rest mass $m$, particle lifetime $t$ and momentum $p$, is (with $v=p/m$)  $x = vt = pt/m$.

Calculating the equivalent decay point under special relativity with $x' = v_{SR}t'$, we have a smaller velocity value $v_{SR} = p/m\gamma_{SR}$ [37] and a larger (time-dilated) decay time $t' = t\gamma_{SR}$, with the two Lorentz factors cancelling ($x'=x$). In this particular calculation, the effect of the time-dilation path-lengthening effect is to compensate for the path-contraction due to special relativity's reduced nominal velocity values, so that the muon's decay position is as it would have been under Newtonian mechanics.

## 11. "SR-similar" aether models

Although this paper is intended to be about recognisably "non-SR" models, we should also mention that there are a range of "Lorentzian" aether models that also incorporate time-dilation effects (see *e.g.* [38]-[40] and many articles in dissident journals). Many of these models only predict small or non-existent deviations from special relativity. Where they agree exactly, the special theory is usually assumed to be preferable because of its reduced number of physical assumptions.

## 12. "Physical" and "interpreted" time dilations

The distinction between "physically-verifiable" time dilation effects and "interpreted" time dilation effects is not always obvious.

In the case of "moving aether" calculations, the "lab-transverse redshift" result is usually overlooked, possibly because it seems unreasonable that a transverse redshift could be detected if the emitter was not "really" ageing more slowly. Since a rectilinearly-moving point-particle only has transverse motion with respect to a point-observer for a vanishingly short period of time, these shifts do not have to be "sustainable", and do not have to be associated with "real" clock-rate differences.

In the "muon" case, time-dilation seems to be an "interpreted" property, whose reality depends on the statement that the speed of light is "really" locked to the observer's own frame – this statement cannot be physically verified without breaking the principle of relativity.

In the case of a relativistic model based on the moving-aether equations (e.g. this author [33][34]), time dilation is more difficult to pin down, as this class of model seems to require a non-Euclidean spacetime in which relative ageing rates can be route-dependent.

### General Summary:

In general:

a) We can test the correctness of the special theory's shift equations, but cannot isolate an unambiguous physical difference in clock rates unless the experiment involves gravitation or acceleration.
   In the absence of these effects, the "time-dilation" results are interpretative.

b) If two physical light-clocks *do* have a verifiable difference in clock rate, then Huyghens' principle applied to this apparent lightspeed differential should give us a description of a gravitational gradient between the two clocks, and the problem can be treated as a gravitational exercise without involving special relativity.

### Exception - "twins" problem

The one possible exception to this rule seems to be the original version of the infamous "twin," "astronaut," or "clock" problem ([43][44], [37] §4.6 pp.125-126), where a traveller coasts away from their twin at a constant $v$ m/s, experiences a sudden abrupt acceleration that reverses their course, and then coasts back to their twin's position at a constant speed of -$v$ m/s.

In special relativity's analysis of the problem, the returning twin shows a final clock-lag equal to the total nominal time-dilation effect accumulated during the constant-velocity stages of its journey. It is difficult to model this outcome gravitationally, since the application of gravitational effects to signals belonging to the slow astronaut's coasting stages can undermine the special theory's calculations [45][46]. If we assume that the sudden acceleration of the traveller produces a shift-inducing gravitational field effect, the characteristics of this abruptly-introduced field are not straightforward [47]. The situation is also difficult to test experimentally.

The favoured GR approach to the problem seems to be to amend the experiment so that the traveller does *not* coast, but experiences a constant acceleration throughout the journey (MTW [22] §6.2-6.6 pp.166-176.).
This, of course, brings us back to a situation where all of the final measurable clock-difference is accumulated while the object is accelerating.

## 13. Checklist

### Inertial motion:

| Model | freq'/freq @$90°_{LAB}$ |
|---|---|
| Stationary absolute aether | 1 |
| Moving absolute aether | 1-vv/cc |
| special relativity | $(1-vv/cc)^{1/2}$ |
| emitter-theory in flat space | 1-vv/cc |
| dragging (extreme) | 1-vv/cc |
| Dragging (intermediate) | 1 *to* 1-vv/cc |
| "relativistic Doppler" equation, applied relativistically [19], no assumption of flat spacetime | $(1-vv/cc)^{1/2}$ |
| "emitter-theory" Doppler equation, applied relativistically [19], no assumption of flat spacetime | 1-vv/cc |
| aether models incorporating time dilation | $(1-v_x v_y/cc)^n$ |
| Aether models incorporating Lorentz time dilation | typically $\sim(1-vv/cc)^{1/2}$ |

**Inertial motion – timeflow:**

| SR Result | Calculable without SR time dilation? |
|---|---|
| Muon track length | Yes |
| "Original" twins problem (combination of inertial and non-inertial motion) | Problematic |

**Non-inertial motion:**

| Test | shift expected without SR? |
|---|---|
| Gravity-shift | Yes |
| Centrifuge test | Yes |
| Haefe-Keating | Yes |
| Orbiting atomic clocks | Yes |
| Rotating object | Yes |
| Thermal atoms | Yes |
| "Constant-g" twins problem | Yes |

## 14. CONCLUSIONS

A small amount of investigation shows that transverse redshifts (where "transverse" means "transverse in the laboratory frame") do seem to appear in most models – of those considered here, only one (flat absolute aether stationary in the observer frame) is not immediately associated with a lab-transverse redshift prediction.

Although it might be considered convenient to dismiss many of these redshift predictions by specifying that the "transverse" detector should be aimed at an angle other than 90° in the laboratory frame, this introduces an additional level of interpretation and theory-dependence into our experiments, and invites confusion about which sets of predictions apply to which experiments.

Since some of these redshift predictions belong to models that predate special relativity and produce *stronger* lab-transverse redshifts than Einstein's special theory, casual statements that "transverse redshifts only appear under special relativity" need to be treated with a certain amount of trepidation.